\documentclass{aa}
\usepackage{graphicx}	% Including figure files
\usepackage{epsfig}
\usepackage{amsmath}	% Advanced maths commands
\usepackage{amssymb}	% Extra maths symbols
\usepackage{verbatim}
\usepackage{color}
\usepackage{txfonts}
\usepackage[version=4]{mhchem}
\usepackage{hyperref}
\usepackage{isotope}

\newcounter{chem}
\newcounter{temp}

\newenvironment{chequation}{%
  \setcounter{temp}{\value{equation}}%
  \setcounter{equation}{\value{chem}}%
}{%
  \setcounter{chem}{\value{equation}}%
  \setcounter{equation}{\value{temp}}%
}

\begin{document}
   \title{Methoxymethanol Formation Starting from CO-Hydrogenation}

   \author{Jiao He\inst{1,2,12}
         \and
		  Mart Simons\inst{3}
		  \and
          Gleb Fedoseev\inst{1,10}
          \and
          Ko-Ju Chuang\inst{1,4}
          \and
          Danna Qasim\inst{1,5}
          \and
          Thanja Lamberts\inst{11,1}
          \and
          Sergio Ioppolo\inst{6}
          \and
          Brett A. McGuire\inst{7,8,9}
          \and
          Herma Cuppen\inst{3}
          \and
          Harold Linnartz\inst{1}
          }

   \institute{
   Laboratory for Astrophysics, Leiden Observatory, Leiden University, PO Box 9513, 2300 RA Leiden, The Netherlands
\and Current address: Max Planck Institute for Astronomy, Königstuhl 17, D-69117 Heidelberg, Germany
\and Institute for Molecules and Materials, Radboud University Nijmegen, Heyendaalseweg 135, 6525 AJ Nijmegen, The Netherlands
\and Laboratory Astrophysics Group of the Max Planck Institute for Astronomy at the Friedrich Schiller University Jena, Institute of Solid State Physics, Helmholtzweg 3, D-07743 Jena, Germany
\and Current address: Astrochemistry Laboratory, NASA Goddard Space Flight Center, Greenbelt, MD 20771, USA
\and School of Electronic Engineering and Computer Science, Queen Mary University of London, Mile End Road, London E1 4NS, UK
\and Department of Chemistry, Massachusetts Institute of Technology, Cambridge, MA 02139, USA
\and National Radio Astronomy Observatory, Charlottesville, VA 22903, USA
\and Harvard-Smithsonian Center for Astrophysics, Cambridge, MA 02138, USA
\and Research Laboratory for Astrochemistry, Ural Federal University, Kuibysheva St. 48, 620026 Ekaterinburg, Russia
\and Leiden Institute of Chemistry, Gorlaeus Laboratories, Leiden University, PO Box 9502, 2300 RA Leiden, The Netherlands
\and \email{he@mpia.de}
             }

\authorrunning{He et al.}
\titlerunning{Methoxymethanol Formation}

  \abstract
   {Methoxymethanol (\ce{CH3OCH2OH}) has been identified through gas-phase signatures in both high- and low-mass star-forming regions. Like several other C-, O- and H-containing COMs (complex organic molecules), this molecule is expected to form upon hydrogen addition and abstraction reactions in CO-rich ice through radical recombination of CO hydrogenation products.  }
   {The goal of this work is to investigate experimentally and theoretically the most likely solid-state methoxymethanol reaction channel – the recombination of \ce{CH2OH} and \ce{CH3O} radicals – for dark interstellar cloud conditions and to compare the formation efficiency with that of other species that were shown to form along the CO-hydrogenation line. Also, an alternative hydrogenation channel starting from methyl formate has been investigated.}
   {Hydrogen atoms and CO or \ce{H2CO} molecules are co-deposited on top of the predeposited H$_2$O ice to mimic the conditions associated with the beginning of ‘rapid’ CO freeze-out. The formation of simple species is monitored \emph{in situ} by infrared spectroscopy. Quadrupole mass spectrometry is used to analyze the gas-phase COM composition following a temperature programmed desorption. Monte Carlo simulations are used for an astrochemical model comparing the methoxymethanol formation efficiency with that of other COMs. }
   {The laboratory identification of methoxymethanol is found to be challenging, in part because of diagnostic limitations, but possibly also because of low formation efficiencies. Nevertheless, unambiguous detection of newly formed methoxymethanol has been possible both in CO+H and \ce{H2CO}+H experiments. The resulting abundance of methoxymethanol with respect to \ce{CH3OH} is about 0.05, which is about 6 times less than the value observed toward NGC 6334I and about 3 times less than the value reported for IRAS 16293B. The results of astrochemical simulations predict a similar value for the methoxymethanol abundance with respect to \ce{CH3OH} factors ranging between 0.06 to 0.03.}
   {We find that methoxymethanol is formed by co-deposition of CO and \ce{H2CO} with H atoms through the recombination of \ce{CH2OH} and \ce{CH3O} radicals. In both the experimental and modeling studies, it is found that the efficiency of this channel alone is not sufficient to explain the observed abundance of methoxymethanol with respect to methanol. The rate of a proposed alternative channel, the direct hydrogenation of methyl formate, is found to be even less efficient.  These results indicate an incomplete knowledge of the reaction network or the presence of alternative solid-state or gas-phase formation mechanisms.  }

   \keywords{astrochemistry -- Methods: laboratory: solid state -- ISM: molecules -- Solid state: volatile}

   \maketitle

   \section{Introduction}
\label{sec:intro}

\begin{figure*}
\centering
\includegraphics[width=10cm]{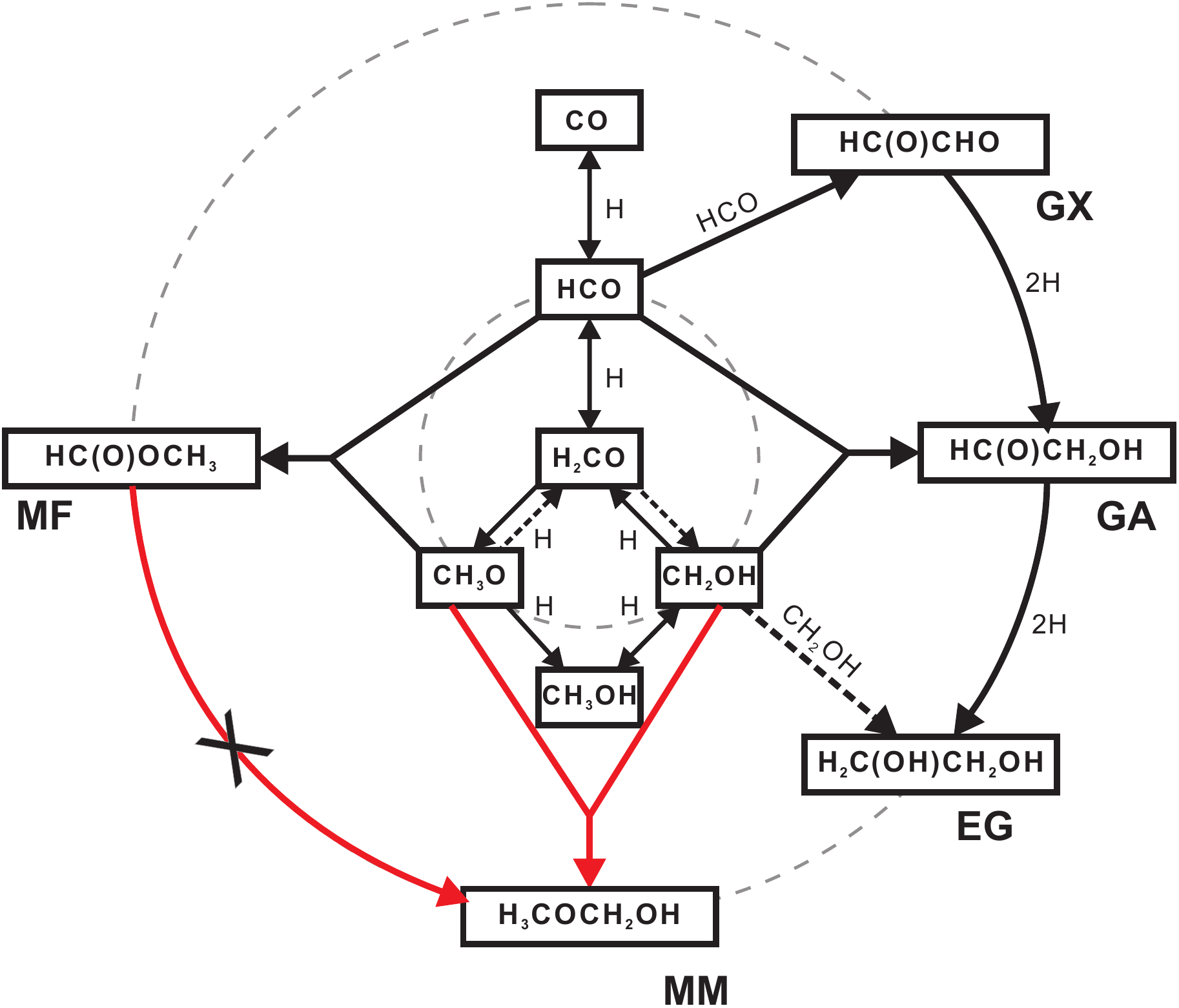}
\caption{Extended COM formation network as obtained from the \ce{CO}, \ce{H2CO}, and \ce{CH3OH} hydrogenation experiments. Solid arrows indicate the reaction pathways confirmed or suggested in this study. Dashed arrows indicate overall less efficient pathways. Updated from Figure 8 in \citet{Chuang2016}. New additions based on this work are shown in red. }
\label{fig:drawing}
\end{figure*}

Ices covering cold dust grains in prestellar cores typically comprise two chemically different layers; a bottom layer dominated by \ce{H2O} and \ce{CO2} (also comprising \ce{NH3} and \ce{CH4}), and a top layer dominated by CO \citep{Pontoppidan2008,Boogert2015}. This two-layer structure reflects different evolutionary phases along the process of star formation; the water-rich phase is dominated by atom addition reactions (O+H, N+H, C+H) and accretion of less volatile species, whereas the formation of the top layer is mainly driven by CO accretion from the gas phase \citep{Herbst2009,Linnartz2015,oberg2016}. In a number of recent studies, this rapid accretion of CO molecules has been shown to act as a starting point leading to the formation of larger, typically C-, O-, N-, and H-containing species. In cold and dark environments, such as prestellar cores, the majority of the chemical processes take place through ``dark chemistry'', i.e., chemistry driven by atom-addition reactions between accreting species reaching thermal equilibrium with the icy grain surface. This is the topic of the present work. In a series of earlier dedicated laboratory studies, it was shown that CO hydrogenation leads to \ce{H2CO} and \ce{CH3OH} formation \citep{Watanabe2002,Fuchs2009}. Addition and abstraction reactions result in the formation of radical intermediates (HCO, \ce{CH3O}, and \ce{CH2OH}) that can recombine with each other or interact with larger (stable) species already present in the ice, to form even more complex species. In this way, it was shown that ongoing hydrogenations, radical-radical, and radical-molecule recombinations can explain the astronomical formation of glycolaldehyde (GA; \ce{CH2OHCHO}), ethylene glycol (EG; \ce{(CH2OH)2}), and to some extent glyoxal (GX; \ce{(HCO)2}) and methyl formate (MF; \ce{CH3OCHO}) \citep{Fedoseev2015,Chuang2016,Chuang2017,Butscher2017,Butscher2019}. This is shown in Figure~\ref{fig:drawing} adapted from \citet{Chuang2016} with a modification (in red) to show the methoxymethanol (MM; \ce{CH3OCH2OH}) formation route under study in this work (note that the formulae are written following the actual molecular geometry). Subsequent laboratory studies showed that following the same principle of consecutive hydrogenation and recombination reactions, larger species, such as glycerol can be formed \citep{Fedoseev2017}. Ultimately, this would provide a pathway towards ribose, a compound of direct biological relevance. 
In the recent past, several studies have been published focusing on solid-state MM formation pathways through energetic processing of \ce{CH3OH} ice or \ce{CH3OH}-containing ices \citep{Maity2015, Paardekooper2016, Schneider2019, Inostroza-Pino2020}. These studies demonstrated that it is possible to form MM under UV light photolysis or upon cosmic rays bombardment. The ionizing irradiation by UV or cosmic ray is less typical in the prestellar phase in which ``non-energetic'' atom addition reactions between thermalized species are most relevant, but may take over at later stages in the star formation process. The involved reactive intermediates, however, are largely similar.

Even though these COMs are expected to form on icy grains, solid-state identifications have not been realized yet in interstellar clouds. Currently, worldwide efforts are underway to collect laboratory COMs ice spectra in support of the upcoming JWST mission set to generate detailed ice maps of the ISM \citep[see e.g.][]{Terwisscha2018,Hudson2019}. It is generally assumed that frozen COMs are liberated from icy dust grains through (non)thermal processes, such as during warm-up of a cloud \citep{Garrod2013} or due to shocks \citep{RequenaTorres2006, Lee2017}. In the gas phase, COMs are typically detected using radio and submillimeter telescopes \citep[see e.g.][and references therein]{Jorgensen2012,Belloche2013,McGuire2018}. Recently, methoxymethanol (MM; \ce{CH3OCH2OH}) has been identified in the gas phase; using ALMA, MM was abundantly detected toward the MM1 core in the high-mass star-forming region NGC 6334I \citep{McGuire2017} at an abundance 34 times less than \ce{CH3OH}, and toward a low-mass star-forming region IRAS 16293-2422 B \citep{Manigand2020}.

The formation of MM through recombination of \ce{CH3O} and \ce{CH2OH} radicals, which are formed by surface CO hydrogenation by accreting H atoms in a typical prestellar setting, has not previously been studied experimentally. The presence of both radicals is expected, as MF, GA, and EG were shown to form through reactions involving these radicals \citep{Chuang2017}. The \ce{CH3O} and \ce{CH2OH} radicals can be formed upon \ce{H2CO} hydrogenation (which is formed by CO hydrogenation) or upon H-atom abstraction from \ce{CH3OH}. CO is, after water, one of the most abundant ice constituents on cold dust grains. \ce{H2CO} ice also has been tentatively identified toward Young Stellar Objects at the 2--7\% level with respect to \ce{H2O} ice \citep{Keane2001, Boogert2015}. \ce{CH3OH} is the largest molecule that has been unambiguously identified in the solid-state on dust grains \citep[][and references therein]{Boogert2015}.

Besides observational and experimental attempts, also the results of modeling studies have been reported in the literature. A three-phase chemical kinetics code \emph{magikal}, first introduced in \citet{Garrod2013}, was used to model the formation of MM, but resulted in an unexpected MM:\ce{CH3OH} ratio of 10$^{-7}$, which is far from the observationally derived value. \citet{McGuire2017} attributed this difference to incomplete reaction networks and unconstrained reaction efficiencies and barriers. New experimental and theoretical studies, as presented here,  provide additional information. 

In the next section, we first describe the performed experiments and their analysis, followed by the experimental results and their interpretations. The description of the experimental work is followed by the numerical simulations utilizing a Kinetic Monto Carlo code presented in \citet{Simons2020}. First, the experimental results are reproduced by the model, then the same parameter space is used to extrapolate these on the astronomical time scales. The resulting \ce{CH3OH}:MM ratios are compared with those observed in the interstellar medium \citep{McGuire2017, Manigand2020}. This work is concluded with a discussion on the differences between the obtained experimental and theoretical findings and the observational values.

\section{Methods}

\subsection{Laboratory experiments}
\label{sec:method-exp}
Experiments are performed using the SURFRESIDE$^3$ apparatus, which has been described in detail in \citet{Ioppolo2013}. Recent updates are available from \citet{Qasim2020}. SURFRESIDE$^3$ is an ultra-high vacuum (UHV) setup with a base pressure of $\sim10^{-10}$ mbar. Ices are grown on a 2.5 cm $\times$ 2.5 cm gold-plated surface that is located at the center of the main reaction chamber. A closed-cycle helium cryostat cools the gold substrate to as low as 8~K. The substrate temperature is monitored using a silicon diode temperature sensor installed at the back of the substrate. A cartridge heater located above the sample can be used to heat the substrate to as high as 450 K. A sapphire thermal switch between the substrate and the cold head of the cryostat enables heating up the substrate to a high temperature without warming up the whole cold head and the surrounding radiation shield. A Lakeshore 340 temperature controller reads and controls the temperature to an accuracy of 0.5 K. Ices on the gold substrate can be monitored both by RAIRS (Reflection Absorption InfraRed Spectroscopy) and by temperature programmed desorption (TPD) using a quadrupole mass spectrometer (QMS). RAIRS is an \emph{in situ} diagnostic tool that allows monitoring ice changes in real-time and without the need for ice heating to realize mass spectrometric detection. However, it suffers from the fact that infrared absorption features of COMs often overlap significantly with each other because of similar functional groups and with the absorption features of other smaller and more abundant species. TPD-QMS is more sensitive --- thermally desorbed species are ionized by an electron source and detected in the gas phase --- but ice destruction is intrinsic to its application. Upon ionization, COMs fragment and show a molecule-specific mass spectra (fragmentation patterns). This provides a further diagnostic tool, in particular when also using isotopically enriched precursors. In this study, we used RAIRS mainly to quantify the amount of CO, formaldehyde (\ce{H2CO}), or MF deposited on the substrate. For this, an FTIR spectrometer is used that covers the 4000-700 cm$^{-1}$ region with a spectral resolution of roughly 1 cm$^{-1}$. Given the rather low abundances of the formed molecules and the spectral overlaps with other species present in the ice, RAIRS was not suited to unambiguously prove MM formation. To identify the molecules that are formed in the ice upon atom bombardment, the ice is heated at a rate of 5~K/minute and TPD-QMS is performed.

The ice samples are grown through co-deposition of molecules and atoms on the precooled substrate. CO gas (Linde, 99.997\%) and \ce{H2CO} vapors or their \ce{^{13}CO} (Sigma-Aldrich, 99 atom \% \ce{^{13}C}), \ce{C^{18}O} isotopologs (Sigma-Aldrich, 95 atom \% \ce{^{18}O}), or $\isotope{H}_2\isotope[13]{CO}$ isotopologs are admitted into the chamber from a pre-pumped dosing line via a variable leak valve. \ce{H2CO} or  $\isotope{H}_2\isotope[13]{CO}$ vapors are prepared in a hot water bath of paraformaldehyde (Sigma-Aldrich, 95\%) or $^{13}$C-paraformaldehyde (Sigma-Aldrich, 99 atom \% $^{13}$C) powder, respectively. MF (Sigma-Aldrich, 99\%) and \ce{H2O} (Milli-Q) vapors are evaporated from their liquid form that went through several freeze-pump-thaw cycles. \ce{H2O} is admitted to the chamber via a separate UHV unit that is connected to the main reaction chamber via a shutter, while MF is admitted through another variable all-metal leak valve.  The amount of CO, \ce{H2CO}, and formed \ce{CH3OH} on the substrate is measured \emph{in situ} from the infrared absorption band area, utilizing setup-specific band strengths reported in \citet{Chuang2018}. For MF, this value is unavailable and is estimated by assuming that the band strength ratio of MF between reflection and transmission modes is similar to that of \ce{H2CO}. The transmission band strength of MF from \citet{Modica2010} ($4.87\times10^{-17}$ cm molecule$^{-1}$, the average values on Si and KBr substrate) is used to extrapolate the MF RAIRS band strength. Hydrogen atoms are produced by a commercially available MicroWave Atom Source (MWAS) with a flux of ~(2-4)$\times 10^{12}$  atoms cm$^{-2}$ s$^{-1}$. The calibration method used to estimate the atomic hydrogen flux is reported in \citet{Ioppolo2013}. Relative COMs formation yields are quantified by analyzing the TPD mass peaks. The mass-to-charge values used for \ce{CH3OH}, MM, GA, and EG are 32, 61, 60, and 33, respectively, while ionization cross-sections are equal to 4.44, 7.16, 6.5, and 7.16 \AA$^2$, which are the same values used in \citep{Chuang2017}. Quantitative analysis of MM is significantly complicated by the lack of literature data for this molecule given its spontaneous decomposition under standard conditions. To the best of our knowledge, no literature values are available for the ionization cross-section of MM. Thus, the value available for EG is adapted for MM because of the structural similarity between the two molecules. The only available full-range mass-spectra is taken from \citet{Johnson1991}. This spectrum reveals a significantly higher fraction of heavy m/z=61 signal with respect to the lower m/z values in comparison to the well-known spectra of GA and EG. Under our experimental conditions, this may result in the systematic underestimation of the amounts of the formed MM. This is important to note, as later on it is concluded that the observational data hint for higher MM abundances. The obtained abundances can be reevaluated if better data become available. The aforementioned spontaneous decomposition of MM on the walls of the setup and the QMS is another factor reducing the observed amounts of MM.

Table~\ref{tab:exp} lists the relevant experiments performed in this study. For each of them, the gold substrate is pre-covered by 5 monolayers (ML,  10$^{15}$ molecules/cm$^2$) of amorphous solid water (ASW) to emulate the water-rich layer on dust grains. The water ice is grown by water vapor deposition when the substrate is at 10 K.
It is well-known that ASW grown under this condition is highly porous, offering a large surface area \citep{Stevenson1999,He2019}.

\subsection{Kinetic Monte Carlo Simulations}
\label{sec:method-modeling}
Ice evolution has been simulated using a KMC algorithm similar to \citet{cuppen2007, cuppen2009}. A similar set-up and network are used as in \citet{Simons2020}. A detailed description of this technique can be found in \citet{Cuppen2013}. The grain surface is described by a 100 × 100 grid of binding sites with periodic boundaries. A random number generator is used to determine the sequence of events from predefined rates for all relevant processes. These processes include deposition of species, hopping between sites on the grid, the reaction between two surface species, desorption from the surface.
This KMC method is used to simulate both experimental and interstellar conditions. Experimental conditions are set to reflect experiments 3 and 5 from Table \ref{tab:exp}. The CO flux has been adjusted to deposit a total of 2.65 ML during the simulation while hydrogen flux is then set to $3\times 10^{12}$ cm$^{-2}$ s$^{-1}$ to be in line with the experiments.
For interstellar conditions, simulations with n(H) = 2.5 and 4.0 cm$^{-3}$ at temperatures of 10 and 12~K are performed. The initial CO gas-phase abundance is set at 10 cm$^{-3}$ for all simulations, and the CO flux is constantly recalculated, taking into account the CO depletion of the gas by CO freeze-out.
The reaction network used is similar to \citet{Simons2020} with the addition of the proposed formation of methoxymethanol: \ce{CH3O} + \ce{CH2OH} $\rightarrow$ \ce{CH3OCH2OH}. This reaction competes with \ce{CH3O} + \ce{CH2OH} $\rightarrow$ \ce{CH3OH} + \ce{H2CO}. This latter reaction has been studied in \citet{Simons2020} and is currently also topic of ongoing experimental work. These reactions are barrierless and their branching ratio is determined to be 0.25:0.65, respectively, with a 0.10 non-reactive branch. This has been done similarly as described in \citet{Lamberts2018}. Furthermore, the rate of the reaction \ce{CH3OH} + H $\rightarrow$ \ce{CH2OH} + \ce{H2} has been updated to $7.22 \times 10^3$ s$^{-1}$ following \citet{cooper2019}.

\begin{table*}[ht]
\caption{Experiments performed in this study. \protect\footnotemark}

\label{tab:exp}
\begin{tabular}{lccccc}
 & Reactants             & T (K) & H dep. (ML) & CO/H$_{2}$CO/MF dep. (ML) & MM:MeOH \\ \hline
Exp. 1 & $^{13}$CO+H      & 10              & 50.0               & 2.7                      & 0.005     \\
Exp. 2 & C$^{18}$O+H      & 10              & 50.0               & 2.1                      & 0.005     \\
Exp. 3 & $^{12}$CO+H      & 10              & 50.0              & 1.5                      & 0.004     \\
 Exp. 4 &  $\isotope{H}_2\isotope[13]{CO}$+H & 10              & 50.0               & 1.7                      & 0.005     \\
 Exp. 5 &  $\isotope{H}_2\isotope[12]{CO}$+H & 10              & 50.0               & 2.0                        & 0.005     \\
 Exp. 6 & $\isotope{H}_2\isotope[12]{CO}$+H & 25              & 50.0               & 2.6                      & -       \\
 Exp. 7 & $^{13}$CO+H      & 25              & 50.0               & 2.0                        & -      \\
 Exp. 8 & MF+H             & 10              & 52.0               & 36.0                        & -
\end{tabular}
\end{table*}
\footnotetext{The amount of CO/H$_2$CO/MF deposited is obtained from the infrared data, while the amount of H deposited is from a separate calibration. After deposition, the ices are heated up at a ramp rate of 5 K/minute to do a TPD. The resulting MeOH:MM ratio in the reaction product is shown. }

\section{Results and Analysis}
\label{sec:results}

\subsection{MM formation through CO and \ce{H2CO} hydrogenation}

\begin{figure}
\centering
\includegraphics[width=0.95\linewidth]{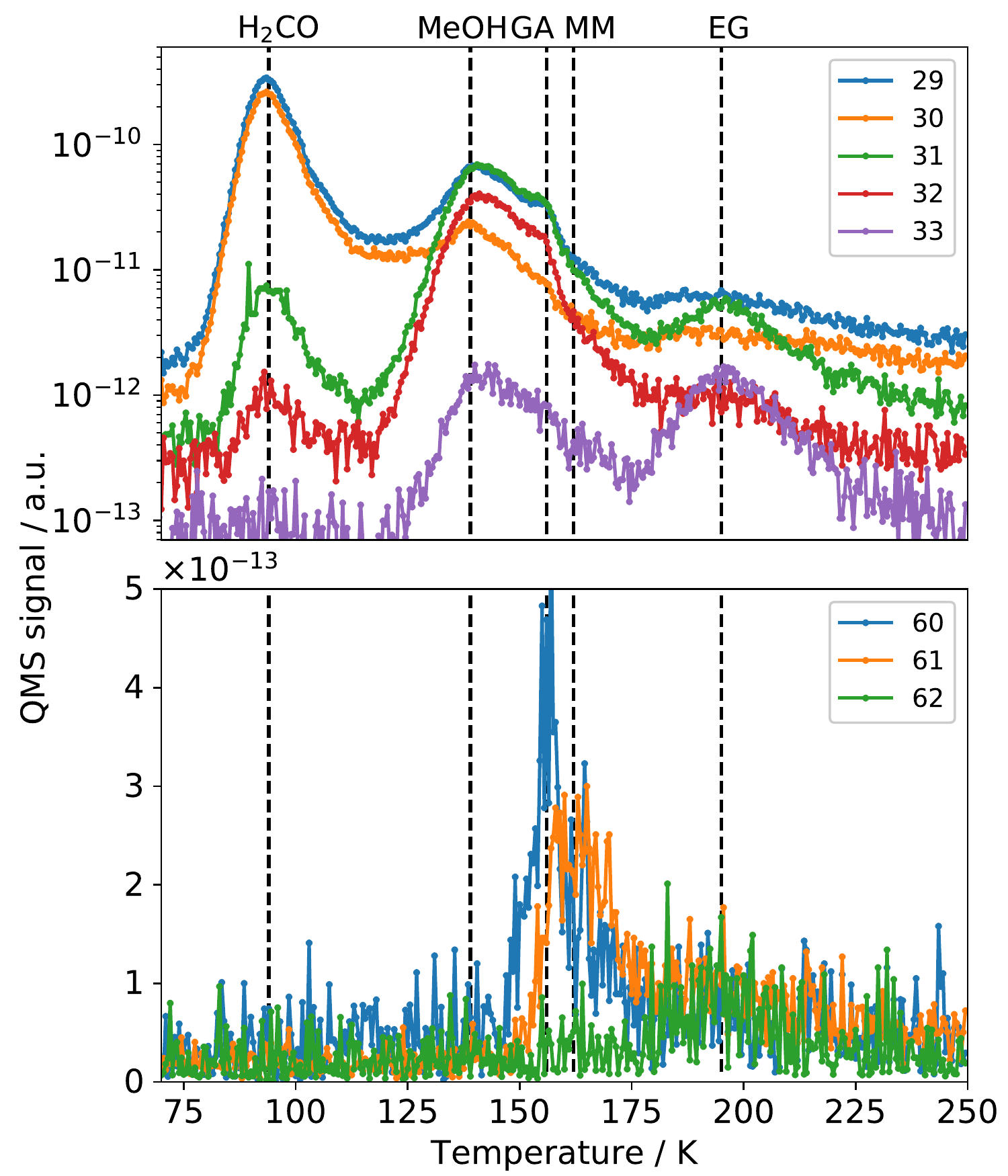}
\caption{QMS spectra of selected masses measured during a 5 K/minute TPD after the co-deposition 2.0 ML of \ce{H2CO} and 50 ML of  H on top of porous amorphous water at 10 K (Exp.1 in Table \ref{tab:exp}).  The assignments of molecular species, based on temperature and masses, are marked in the figure. \ce{CH3OH} is abbreviated to MeOH.}
      \label{fig:191218}
\end{figure}

Figure~\ref{fig:191218} shows the TPD spectra during a temperature ramp of 5 K/minute after a \ce{H2CO} and H-atom co-deposition experiment at 10~K (Exp. 5). The spectra show clear evidence for the presence of initial \ce{H2CO} precursor and its hydrogenation product, \ce{CH3OH}, peaking at temperatures of 95 (m/z = 29, 30)  and 139 K (m/z = 29-33), respectively. The figure also shows that larger COMs are formed, such as GA and EG that can be unambiguously detected at 156 and 195 K by their molecular mass signals (m/z = 60 for GA, and m/z = 62 for EG) and shared mass fragments (m/z = 29, 30, 31, 32, and 33) \citep{Chuang2016}. As the formation of these 2-carbon COMs involves the same radicals (\ce{CH3O} and \ce{CH2OH}) needed to form methoxymethanol, it makes sense to search for MM mass signatures as well, which turned out to be quite a challenge.

\begin{figure}
  \centering
  \includegraphics[width=0.95\linewidth]{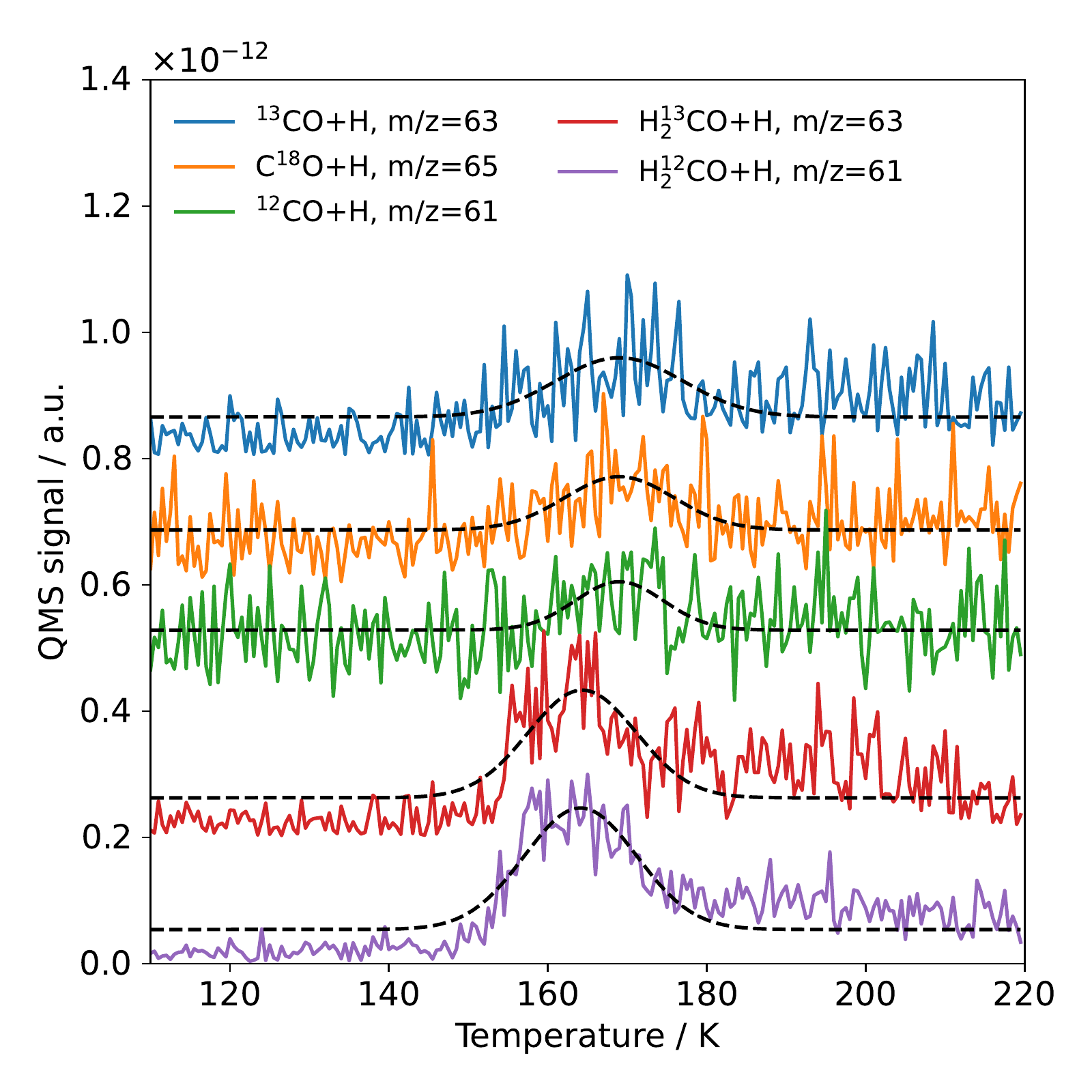}
  \caption{QMS spectra of methoxymethanol during the TPD after co-deposition of H with $^{13}$CO (Exp. 1), C$^{18}$O (Exp. 2), CO (Exp. 3), H$_2^{13}$CO (Exp. 4), or \ce{H2CO} (Exp. 5). Spectra are offset for clarity. A Gaussian fitting for each curve is shown in a black dashed line. }
        \label{fig:MM_new}
  \end{figure}

The possible formation of MM from UV irradiation or electron impact of methanol-containing ices has been studied previously by several groups. All of them found TPD peaks for an m/z= 61 (or shifted if isotope-labeled precursors were used) between 155 and 170 K and attributed this peak to MM. Table \ref{tab:MM_lab_works} summarizes these studies. According to the GC/MS spectrum of MM reported in \citet{Johnson1991}, m/z = 61 is one of the strongest MM signals, while the dissociative ionization peak at m/z = 60 and the peak of the undissociated ion at m/z = 62 are only minor contributors to the total ion current. This was also verified by the experiments in \citet{Boamah2014} and \citet{Sullivan2016}. In our work, we also find that m/z = 61 peaks at about 162 K, which differs by 6 K from GA in terms of desorption temperature. This detection is consistent with prior identification of MM in TPD, both in temperature and mass-to-charge values. In all other CO+H or \ce{H2CO}+H experiments performed at 10 K, a peak at m/z = 61 amu is found around 162 K, but this signal disappears in experiments where the ice was deposited at 25 K. This is expected because at 25 K H-atoms have a very short residence time on the surface and therefore do not drive efficient chemistry, as illustrated earlier for the CO $\leftrightarrow$ \ce{H2CO} $\leftrightarrow$ \ce{CH3OH} reaction sequence \citep{Watanabe2002, Fuchs2009, Chuang2017}. Further support for the assignment of m/z = 61 at 162 K peak to MM comes from TPD spectra obtained when using different isotope precursors. In Figure~\ref{fig:MM_new}, the corresponding TPDs are shown (top-down) after hydrogenation of \ce{^{13}CO}, \ce{C^{18}O}, \ce{^{12}CO}, $\isotope{H}_2\isotope[13]{CO}$, and $\isotope{H}_2\isotope[12]{CO}$ ices. 
It can also be seen that when starting from CO, the desorption peak temperature of MM is slightly higher, which is probably due to the lower coverage of MM on the surface. The use of isotopes results in the corresponding shifts for the mass-to-charge value of m/z = 61 peak of MM to account for the extra mass of $^{13}$C or $^{18}$O label in the obtained ion. It should be noted that the experiments starting from \ce{H2CO}+H have a higher absolute yield of MM than those starting from CO+H. Exps. 1, 2 and 3 (starting from CO) the TPD signals are weak, for Exps. 4 and 5 (starting from \ce{H2CO}) the signals are clear. For all of them, the relative yield of MM to \ce{CH3OH} are comparable \ref{tab:exp}. The higher yield of MM when starting from \ce{H2CO} is in line with the formation of MM along the \ce{H2CO} $\leftrightarrow$ \ce{CH3OH} part of the hydrogenation network. Two extra hydrogenation steps are required to reach MM from CO compared to \ce{H2CO}. Given these and previous findings, we conclude that MM is also formed in the CO hydrogenation chain, although with a lower abundance compared to those of GA and EG.

\begin{table*}
\caption{Previous laboratory works related to MM formation in the solid state}
\label{tab:MM_lab_works}
\centering
\begin{tabular}{lccc}
\hline\hline
 Prior works &  Ice sample   &  Irradiation & Basis of identification  \\
\hline
\citet{Harris1995}  &  \ce{CH3OH} & $\le 55$ eV e$^{-}$ &   TPD,  61/33/45 amu at 170 K \\
\citet{Boamah2014} &  \ce{CH3OH} &  7 and 20 eV e$^{-}$  & TPD,  61/62 amu at 165 K, isotope \\
\citet{Boyer2014} &  \ce{CH3OH} &  20 and 1k eV e$^{-}$  & TPD,  61 amu at $\sim$170 K \\
\citet{Maity2015}  &  CO:\ce{CH3OH} &  5k eV e$^{-}$    & TPD,  61 amu at 160--170 K, isotopes \\
\citet{Sullivan2016} &  \ce{CH3OH} &  20 and 1k eV e$^{-}$ & TPD,  61/62 amu at $\sim$165 K \\
\citet{Paardekooper2016} &  \ce{CH3OH}&  10.2--7.2 eV UV & laser desorption TOF-MS, 150--175 K, tentative\\
\citet{Schneider2019} &  \ce{CH3OH}  &  6.7--7.4 eV UV   & TPD,  61 amu at 155 K, isotopes\\
\hline
\end{tabular}
\end{table*}

\subsection{MM formation through MF hydrogenation}

As shown in Figure~\ref{fig:drawing}, it is possible, in principle, to form MM through the direct hydrogenation of methyl formate. This route
\begin{chequation}
  \begin{align}
\ce{CH3OCHO} & \xrightarrow{\text{2H}} \ce{CH3OCH2OH}   \label{r:mf_mm}
\end{align}
\end{chequation}
 was also proposed by \citet{McGuire2017}. MF is an abundant two-carbon COM observed in hot cores, hot corinos, and comets \citep[][and references therein]{Taquet2017, Bacmann2012}. It is the most abundant isomer of GA. Although it has not been detected yet in the solid-state on dust grains, it is likely one of the COMs embedded in the ice mantle \citep{Garrod2013}. New IR data of MF ice have become available recently \citep{Terwisscha2021} that allow searching for solid MF in space.

\begin{figure}
\centering
\includegraphics[width=0.95\linewidth]{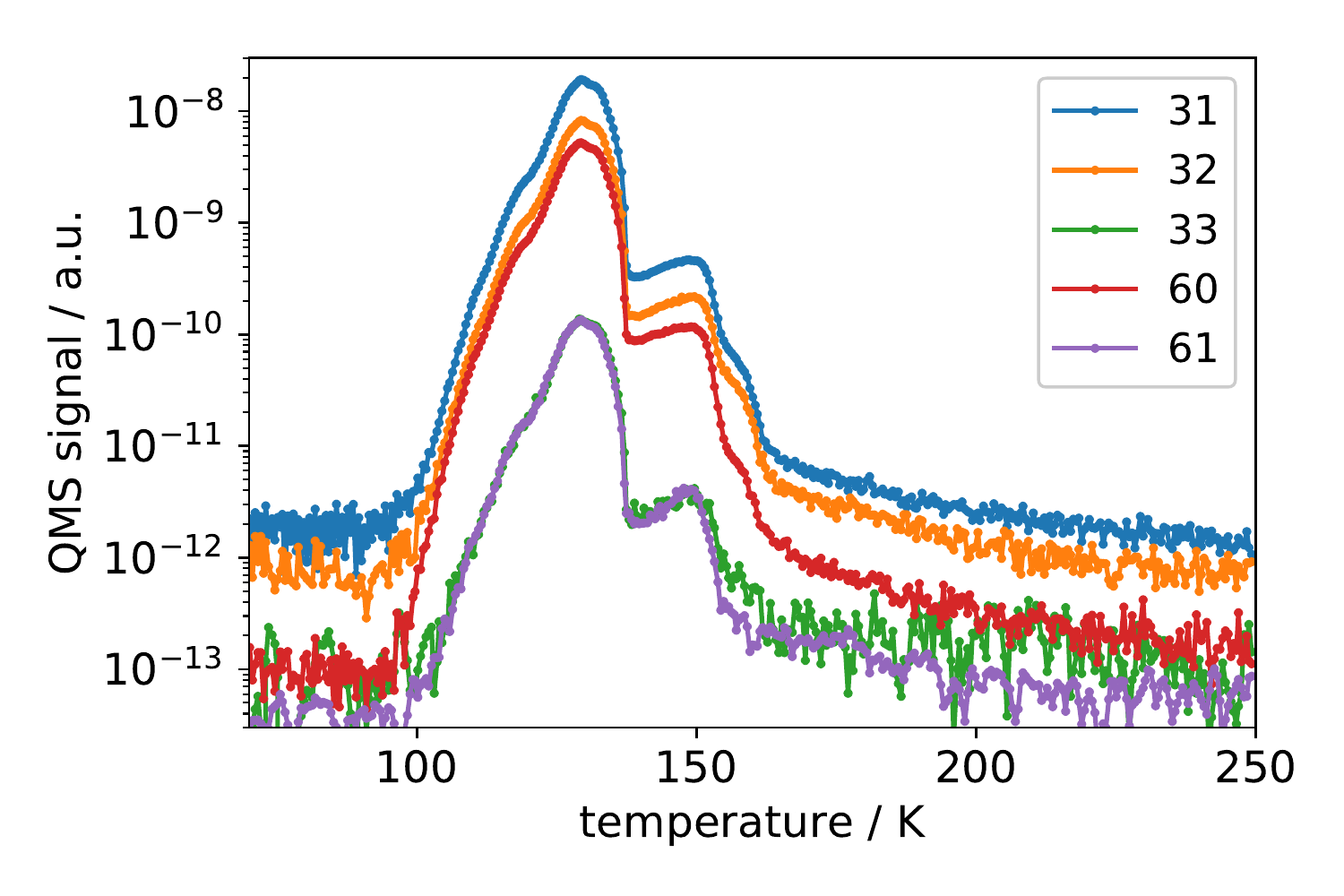}
\caption{TPD spectra of 36.0 ML methyl formate and 52.0 ML H co-deposited at 10 K and then heated up at 6 K/minute. The mass corresponding to each curve is shown in the inset. The red vertical dashed lines marks the temperature 162 K at which the desorption of MM is expected. }
\label{fig:mf_h}
\end{figure}

\begin{figure}
\centering
\includegraphics[width=0.95\linewidth]{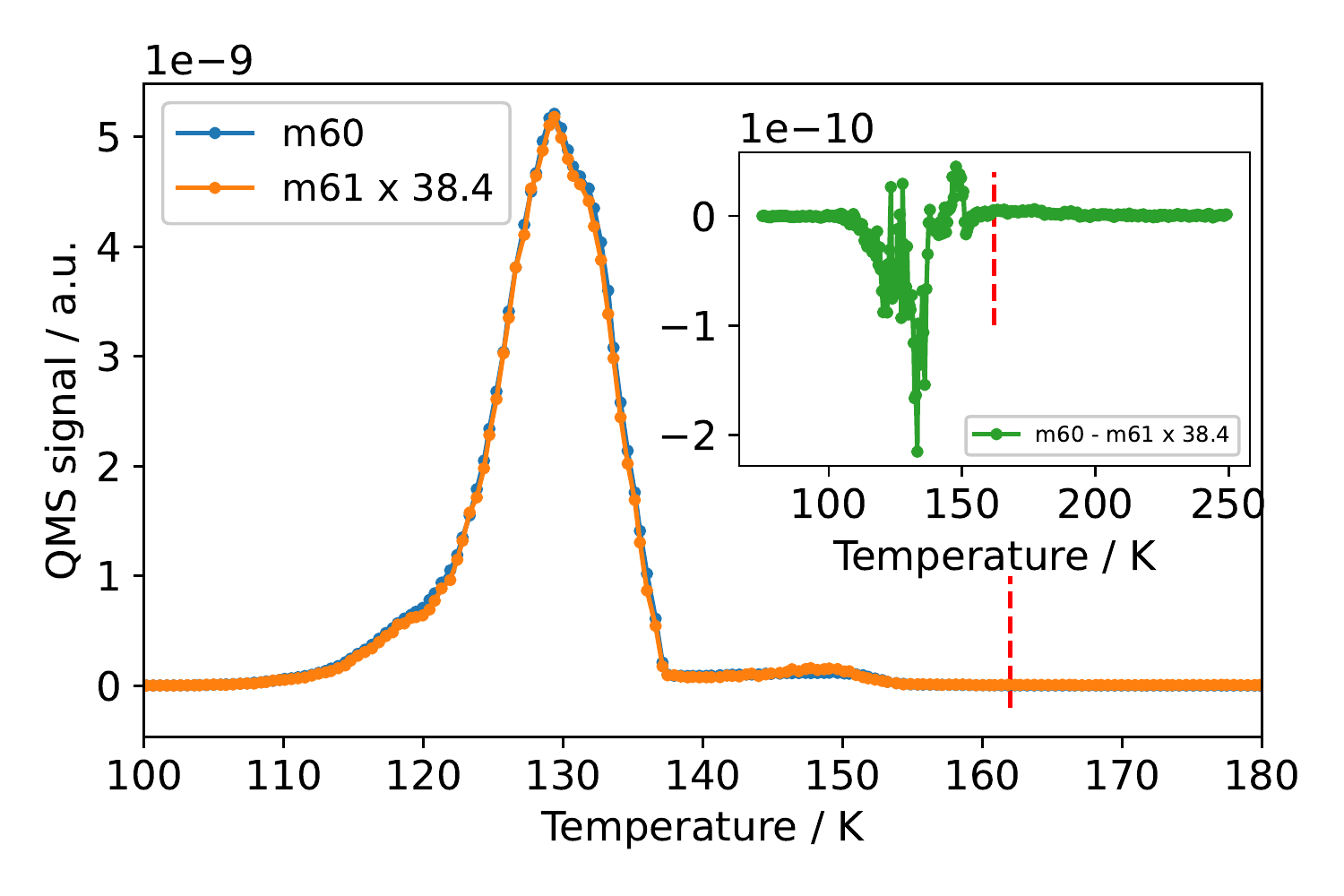}
\caption{A comparison between the QMS signal of m/z=60 and 61 (multiplied by 38.4 times) taken from the corresponding curves in Figure~\ref{fig:mf_h}. The baselines for both masses are subtracted. The inset shows the difference between the two traces. }
\label{fig:mf_h_diff}
\end{figure}

 In this study, we have tested experimentally the MF+H reaction pathway (Exp. 8). With the gold substrate at 10 K, 36.0 ML MF and 52.0 ML H are co-deposited onto the substrate over three hours. Subsequently, the ice is warmed up from 10 to 250 K at a ramp rate of 6 K/minute to do a TPD. The TPD spectra are presented in Figure \ref{fig:mf_h}. The main desorption peak of MF is located at ~135 K. At ~148 K, there is another small peak, which is 15-20 K lower than the desorption peak of MM. Following the work by \citet{Zahidi1994}, we attribute this peak to the desorption of MF in the submonolayer. The interpretation of a possible MM signature is not a priori clear. However, based on the mass fragmentation pattern of MF from the NIST database, the m60/m61 ratio should be 38.4. If there is a significant amount of MM in the ice, we would expect to find a lower m60/m61 ratio at the MM desorption temperature ~162 K, given the extra contribution originating from MM. To better see whether there is any sign of MM at $\sim$162 K, in Figure~\ref{fig:mf_h_diff} we plot the m60 and 38.4 x m61 signal versus temperature. It is clear that these two curves are almost identical and no evidence of MM desorption can be seen at 162 K. To further verify this, in the inset, we plot the difference m60 - 38.4 x m61, and still see no desorption peak of MM at 162 K. The small difference between $\sim$110 and $\sim$155 K could be at least partly due to the small time offset between the two mass channels, as the QMS can only scan one mass channel after another. Therefore, we conclude that there is insignificant MM formation for the investigated settings and available QMS sensitivity. This is in agreement with a recent study by \citet{Krim2018} who combined laboratory experiments with theoretical calculations and found that hydrogen addition to MF has a non-trivial activation energy barrier of 32.7 kJ mol$^{-1}$, likely rendering this route inefficient under interstellar relevant conditions. This finding also agrees with \citet{Alvarez2018} who found that the rate constant of H+MF is several orders of magnitude lower than that of H+GA based on instanton theory calculations.

\subsection{Astrochemical modeling}
\label{sec:model}
\subsubsection{Simulated Experiments}
\label{sec:sim_exp}
\begin{figure}
\centering
\includegraphics[width=0.95\linewidth]{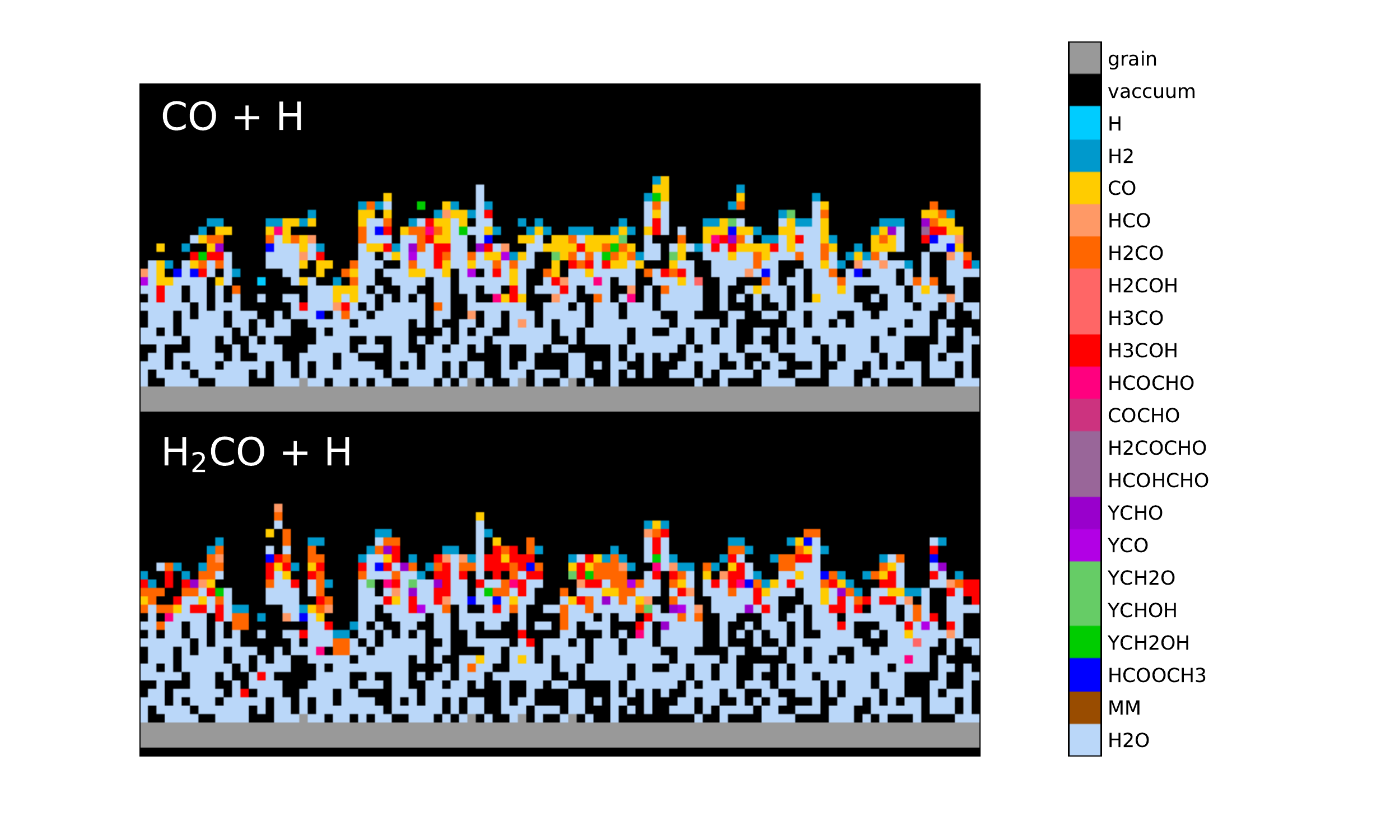}
\caption{Cross-sections of the grain mantle after 6 hours of simulated experiments. }
\label{fig:simons_fig1}
\end{figure}

We have used Monte Carlo simulations to obtain molecular details on the exact formation route of MM. To obtain an accurate surface, 10 monolayers of water were deposited onto a bare surface before the co-deposition simulation. This water layer is porous and yields an approximate surface area of 2 ML at the solid-vacuum interface. Figure~\ref{fig:simons_fig1} shows a cross-section of this water layer in light blue on top of the gray-colored grain.

\begin{figure}
\centering
\includegraphics[width=0.95\linewidth]{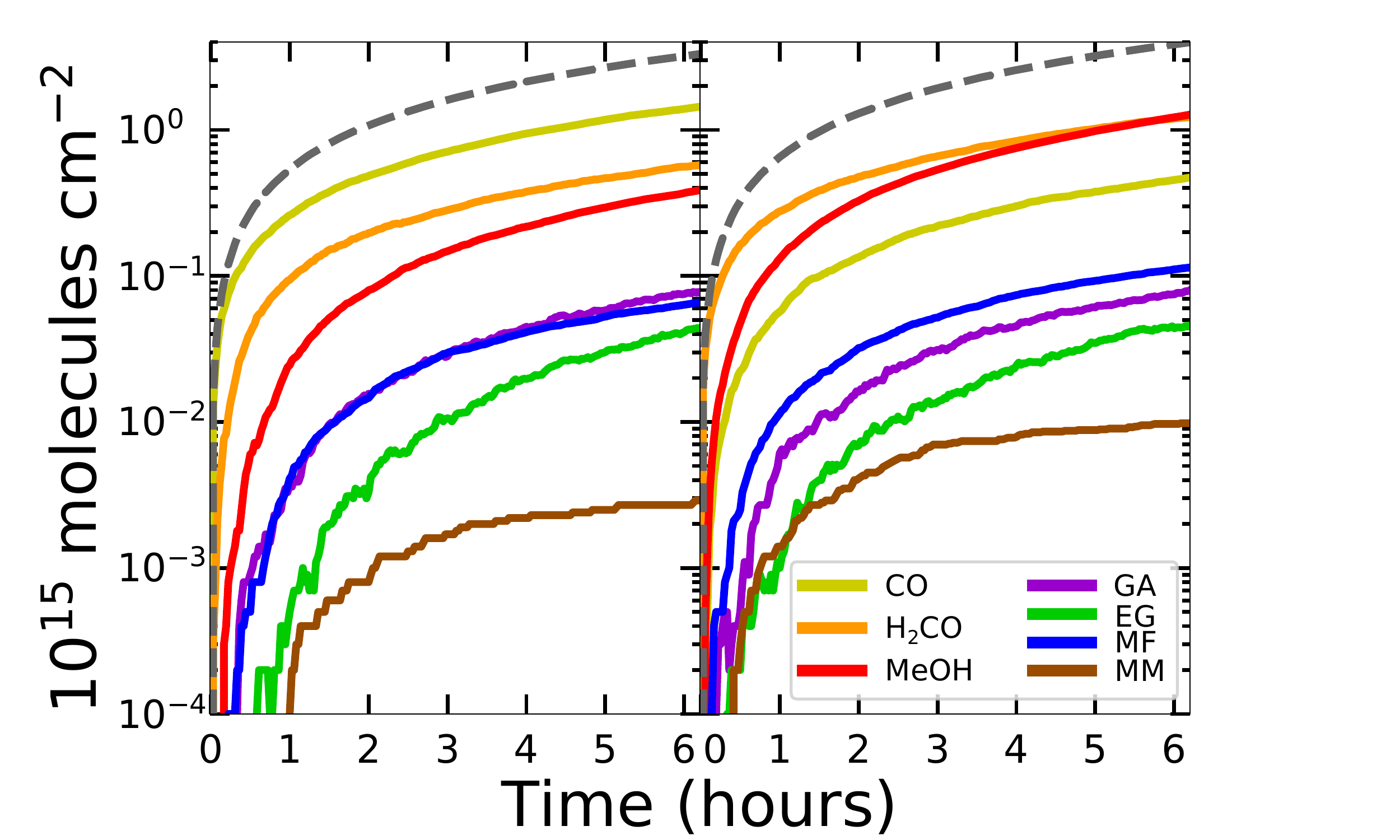}
\caption{Surface abundance of stable species during benchmark simulations of CO+H (left) and \ce{H2CO}+H (right) co-deposition. These are simulations of Exp. 3 and 5 as listed in Table \ref{tab:exp}.   The dashed line is the total amount of C deposited. }
\label{fig:simons_fig2}
\end{figure}

\begin{figure}
\centering
\includegraphics[width=0.95\linewidth]{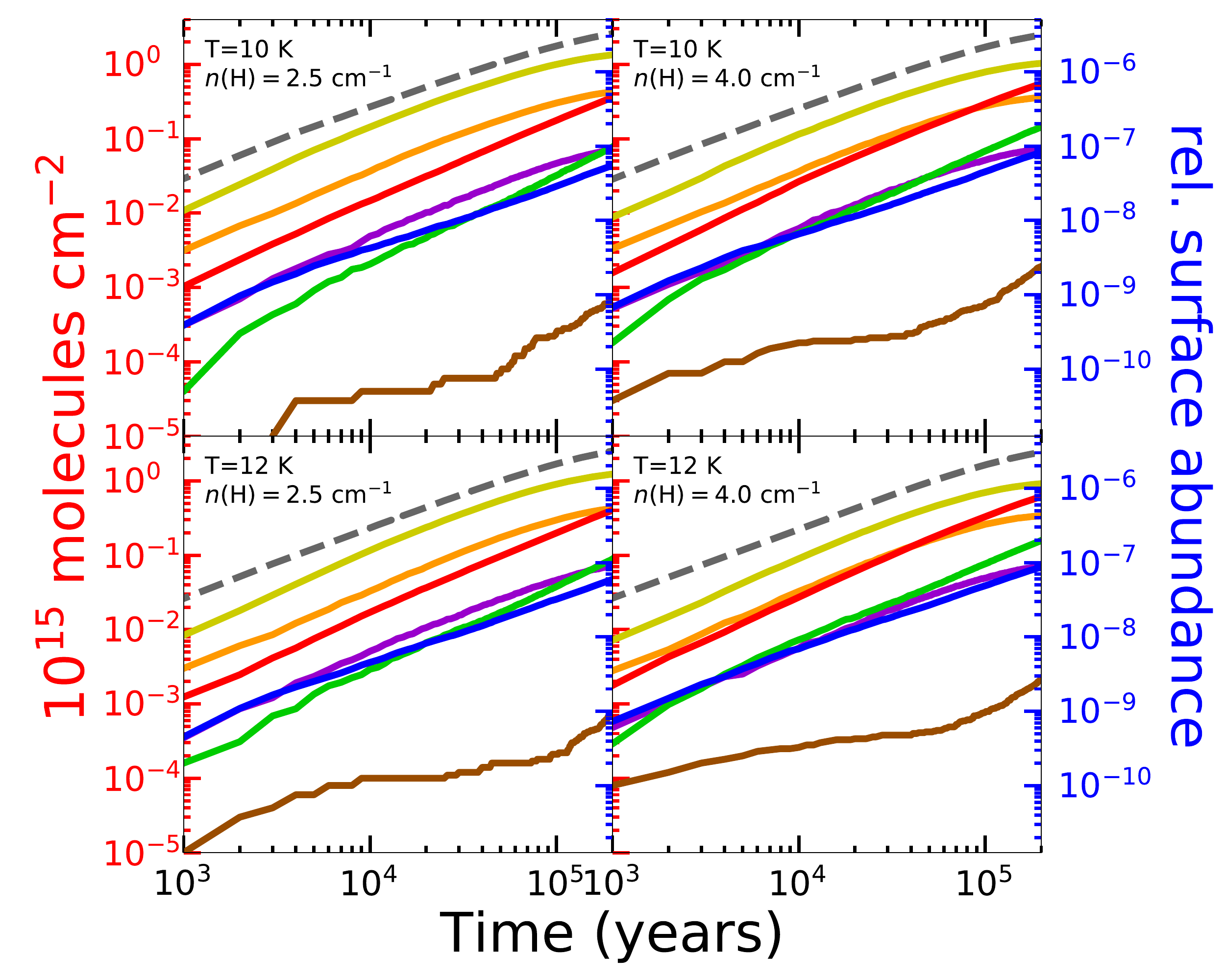}
\caption{Simulated time evolution of the grain surface under interstellar conditions. The initial CO abundance is equal to 10.0 cm$^{-3}$ in all simulations. Surface abundance is given with respect to $n_H$. Coloring is consistent with Fig. \ref{fig:simons_fig2}. }
\label{fig:simons_fig3}
\end{figure}

For the experimental simulations, the binding energy of the H and CO was increased by 20 percent with respect to work in \citet{Simons2020} to reflect the change in ice substrate from CO-rich to \ce{H2O}-rich. Figure~\ref{fig:simons_fig2} shows the surface abundance of stable species during the simulation. It can be seen that, in time, hydrogenation reactions become more efficient than other radical-radical reactions that lead to MF (blue) and MM (brown). This is a result of the first monolayer forming. When approaching a full monolayer of deposited CO, the probability of a newly deposited hydrogen atom to find a species increases greatly, because there are more molecules to land on top of. The efficiency of hydrogenation is hereby increased, as less hydrogen diffusion is required. This in turn decreases the probability of two non-hydrogen radicals being in close proximity since most radicals are quickly hydrogenated. This effect is more prevalent for the formation of MM since the formation of \ce{CH2OH} is already quite inefficient. This is reflected by an almost fully flattened curve of MM towards the end of the simulation.

\begin{table*}[]
  \caption{Coverage of stable species at the end of 6 hours of simulated co-deposition in experimental conditions.}
  \label{tab:model}
  \begin{tabular}{ll|ccc|cccc|c}
  \hline\hline
  T  & n(H) & \multicolumn{3}{c|}{Coverage (MLs)} & \multicolumn{4}{c|}{Ratio w.r.t. MeOH} & MM:COMs \\ 
     (K)  &     (cm$^{-3}$) & CO         & \ce{H2CO}      & MeOH      & GA      & EG      & MF      & MM      &         \\ \hline
  10 & 2.5  & 13.420     & 4.252     & 3.604     & 0.200   & 0.215   & 0.121   & 0.002   & 0.004   \\
  10 & 4.0  & 10.316     & 3.588     & 5.478     & 0.134   & 0.262   & 0.119   & 0.003   & 0.004   \\
  12 & 2.5  & 12.294     & 4.239     & 4.022     & 0.180   & 0.221   & 0.118   & 0.002   & 0.003   \\
  12 & 4.0  & 9.155      & 3.368     & 6.136     & 0.115   & 0.257   & 0.112   & 0.003   & 0.007   \\ \hline
  \end{tabular}
  \end{table*}

Table~\ref{tab:model} lists the surface coverage of the non-complex species at the end of the six-hour simulation. Generally, we obtain slightly more hydrogenated species than earlier works because of the higher n(H):n(CO) ratio used in these simulations.  The absolute error in coverage is within a factor 2 uncertainty of the IR data of the experiments. Comparing the amount of COMs with the QMS data shows that the simulations yield much more methyl formate than the experiments. This is probably the result of missing destruction routes of large molecules in the reaction network. The simulations also give a slightly higher surface abundance of MM with respect to Table \ref{tab:exp}. The QMS data are however expected to be a lower limit with a fairly high uncertainty, so the increase of approximately 30 percent is well within a reasonable margin. Overall, Table~\ref{tab:model} shows that MM constitutes only a small fraction of all formed COMs. This is in line with experimental results reported earlier.

Finally, the co-deposition of \ce{H2CO} and H yields a higher abundance of MF and MM compared to CO + H co-deposition. This can simply be attributed to the fact that fewer hydrogenation steps are required to yield its reactants \ce{CH3O} and \ce{CH2OH}. COMs that require HCO-dimerization (GA and EG) are not affected.

\subsubsection{Simulated Interstellar conditions}
Next, we have performed simulations under interstellar conditions to obtain information that helps to decide where to search for MM in astronomical observations. Firstly, we conclude that MM is formed under all relevant conditions, albeit not in high abundance. The formation of MM is more efficient at 12 K than at 10 K since an increased temperature allows for more efficient surface diffusion of hydrogen. Based on previous studies (\citep{Simons2020}), we expect this trend to continue onto higher temperatures until hydrogen will desorp. Simulations with a hydrogen abundance of 4.0 cm$^{-3}$ yield more MM than those with a lower abundance of 2.5 cm$^{-3}$. This is trivial since a higher hydrogen abundance directly increases the efficiency of hydrogenation reactions required for the formation of MM. This effect is more prevalent in the submonolayer regimes for reasons explained in Section~\ref{sec:sim_exp}, and is also reflected in the surface abundance of MM in Figure ~\ref{fig:simons_fig3}.

Contrary to simulations done in experimental conditions, we now see an increase in MM formation in the late stages of the simulation. This is the result of the CO freeze-out. As more CO is depleted from the gas phase, the n(H):n(CO) ratio of the gas is increased. We know from previous studies that a high n(H):n(CO) ratio increases methanol formation, with methanol being the dominant species at n(H):n(CO) ratios of six and above \citep{Simons2020}. Since \ce{CH2OH} is the product of methanol dehydrogenation and \ce{CH3O} is the precursor of methanol, the formation of these species has a similar dependence on the n(H):n(CO) ratio as methanol. These species together can form MM when formed in close proximity on the grain. Thus, the formation of MM is more efficient at a high n(H):n(CO) ratio, which occurs when CO freeze-out has progressed sufficiently.

\section{Astrophysical Implications}
\label{sec:astro}
In this study, we conclude that MM can be formed in the CO-rich layer of the ice mantle following the hydrogenation of CO or \ce{H2CO} through recombining \ce{CH3O} and \ce{CH2OH}. The direct hydrogenation of MF is not confirmed as an efficient pathway. These findings complete the network introduced in \citet{Chuang2016}. It appears that when starting from \ce{H2CO}, the MM yield is higher. This confirms, not surprisingly, that the formation of MM is more efficient when a significant fraction of the CO has already been hydrogenated to \ce{H2CO} and \ce{CH3OH}. The work presented here indicates that the overall MM formation is (substantially) less efficient than for GA or EG. In the observation toward NGC 6334I, a MM:MeOH abundance ratio of 0.03 was derived \citet{McGuire2017}. Here we calculate this ratio for the studied pathways following a similar procedure as presented in \citet{Chuang2017}. We use m/z values of 32 and 61 amu to quantify the methanol and MM yields, respectively. As the MM ionization cross-section for 70 eV electron impact is unavailable in the literature, we assume a similar cross-section value as for EG, as EG is an isomer of MM. The error introduced by this assumption is estimated to be less than 15\%, based on the known cross-section values for MF, GA, and EG. In the right-most column of Table~\ref{tab:exp}, the calculated MM:MeOH ratios are listed. The values are about 0.004--0.005 for those experiments that identified MM. This means that the solid-state formation efficiency in the laboratory is roughly a factor of six below the ratio detected in the gas phase in space by McGuire and coworkers. This can have several reasons. The calculated ratio may vary with other parameters, such as the surface temperature. As stated before, assumptions concerning the ionization cross-section of MM may result in an underestimation of the amounts formed. The experiments may also underestimate the yield of MM because of unknown destruction mechanisms of MM in the chamber. Finally, as long as the mechanism transferring solid-state COMs into the gas phase is poorly understood, one has to be careful comparing abundance ratios in both phases \citep{Bertin2016,Ligterink2018,Chuang2018b}. Nonetheless, our experimental result suggests that hydrogenation of CO and/or \ce{H2CO} in the outer layer of the ice mantle can at least partially account for the formation of MM. ``Energetic processing'' of methanol-containing ice mixtures, which has been proposed by several groups to be an efficient formation route of MM (see Table~\ref{tab:MM_lab_works}), likely plays an important role. The \ce{CH3O} and \ce{CH2OH} radicals produced by photolysis of \ce{CH3OH} in CO-rich ice recombine to form MM. As the diffusion of radicals in bulk ice is typically inefficient at $\sim$10~K, non-diffusive mechanisms are likely involved in the reactions. \citet{Jin2020} proposed that radicals could be produced right next to each other, therefore chemical reaction occurs without the need to overcome a diffusion energy barrier. \citet{Mullikin2021} found that the inclusion of nonthermal reactions and suprathermal species better reproduces the low-temperature solid-phase photoprocessing in ices of dense cores such as TMC-1. In CO-rich ice, the transition from amorphous to polycrystalline phase is accompanied by the segregation and clustering of minor components in the ice, including radicals such as \ce{CH3O} and \ce{CH2OH} \citep{he2021}. This is also a likely mechanism to explain the formation of MM without involving diffusion. Most recently, \citet{Ishibashi2021} used Cs$^+$ ion pickup method to study the photolysis of \ce{CH3OH} on top of water surface and found the OH radical from the dissociation of water might be important for the formation of MM. The H addition reactions in the current study combined with all these alternative mechanisms present a more complete picture of the formation of MM.

From simulations in interstellar conditions, we conclude that MM is most likely found in pre-stellar objects with a high n(H):n(CO) gas-phase ratio. These are predominantly late-stage dark clouds or early-stage hot cores with a progressed CO freeze-out. IRAS 16293–2422 A is such an object which has been observed in the ALMA-PILS survey \citep{Manigand2020}. An observation of MM has been reported in the survey. However, as this is a gas-phase observation, we cannot make a direct comparison with this study for the formation route, since many of the precursors such as \ce{H2CO} are known to have rich gas-phase chemistry as well. Future observations with JWST can elaborate more accurately on the grain-surface chemistry, even though it was found recently, that it is exceptionally hard to pre-deposit pure MM ice and record high-resolution IR spectra as for other frozen COMs \citep[][]{Rachid2021}.  Solid MM, clearly, is much harder to tackle than its gas phase equivalent.

\section{Conclusions}
\label{sec:conclusions}
In this work, we carried out laboratory experiments under prestellar core relevant conditions and modeling to study the astrochemical relevance of hydrogenation reactions of CO-ice, \ce{H2CO}-ice, and \ce{CH3OCHO}-ice as pathways towards methoxymethanol. The reactions between atomic hydrogen and CO and \ce{H2CO}, followed by \ce{CH3O} and \ce{CH2OH} radical recombinations (as indicated in \citet{Chuang2017}) are concluded to produce methoxymethanol, but generally with a (substantially) lower efficiency than other products in the CO ice hydrogenation chain. Even though the overall ratio of methoxymethanol formed in the solid-state is lower than found in observational gas-phase studies, we conclude that "non-energetic" solid-state processes in the CO-rich layer on the ice mantle take place and at least partially will contribute to the formation of methoxymethanol in dense clouds. 

\section{Ackowledgement}
This work has been financially supported through an NWO grant within the framework of the Dutch Astrochemistry Network II. We thank Julie Korsmeyer for her technical assistance. G.F. acknowledges financial support from the Russian Ministry of Science and Higher Education via the State Assignment Contract. FEUZ-2020-0038. S.I. acknowledges support from the Royal Society.
%\input{methoxymethanol.bbl}
%\bibliographystyle{aa}
%\bibliography{mm}

\end{document}